\documentclass[floatfix,twocolumn,rmp]{revtex4}

\usepackage{graphicx,amsmath,amssymb}

\begin{document}

\title{Dynamic scaling regimes of collective-decision making}

\author{Andreas Gr\"{o}nlund}
\affiliation{Department of Physics, Ume{\aa} University,
901 87 Ume{\aa}, Sweden}
\affiliation{Ume{\aa} Plant Science Centre, Department of Plant Physiology,
University of Ume{\aa}, 901 87 Ume{\aa}, Sweden}

\author{Petter Holme}
\affiliation{Department of Computer Science, University of New Mexico,
  Albuquerque, NM 87131, U.S.A.}
\affiliation{Department of Computational Biology, School of Computer Science
  and Communication, Royal
  Institute of Technology, 100 44 Stockholm, Sweden}

\author{Petter Minnhagen}
\affiliation{Department of Physics, Ume{\aa} University,
901 87 Ume{\aa}, Sweden}

\begin{abstract}
We investigate a social system of agents faced with a binary
choice. We assume there is a correct, or beneficial, outcome of this
choice. Furthermore, we assume agents are influenced by others in making 
their decision, and that the agents can obtain information that may guide 
them towards making a correct decision. 
The dynamic model we propose is of nonequilibrium type, converging to
a final decision. We run it on random graphs and scale-free networks.
On random graphs, we find two distinct regions in terms of the
\textit{finalizing time}---the time until all agents have
finalized their decisions.
On scale-free networks on the other hand, there does not seem to be any such
distinct scaling regions.
\end{abstract}

\maketitle

\section{Introduction}

Designing simple, mechanistic models of social phenomena has been an
off-the-mainstream theme in economics and sociology literature for
several decades~\cite{schelling}. Such models turn out to be very apt
for the methods of statistical physics. So as the statistical physics
community gets increasingly interdisciplinary it is not surprising
that there is a growing interest in ``sociophysics.''  One of the most
common topics in this area is the spread of
opinions~\cite{sznajd,deuff:opi,sood:vot,our:vot} or
fads~\cite{watts:fad,our:fad2}. In models of these phenomena, the
agents do not usually actively use information to optimize their behavior; the spread
takes place more or less like an infection process. Another example of
social systems studied by physicists is spatial, or networked,
games (see
Refs.~\cite{lindgrennordahl,ebelbornholdt:coevo,our:realpd} and
references therein). In that case the agents actively try to
maximize their individual gain. Between these two extremes is the
problem of the present paper---collective decision making~\cite{zucker,fore:marsili}.
This is the situation when a population, through individual
action and social interaction, has to make up their mind about a specific question.
This problem differs from the opinion dynamics in that an individual can do
better or worse. It also differs from evolutionary games in that
there is no conflict built into the problem---the success of one agent
is never the harm of others, or vice versa.

In this paper we model collective decisions as one at the same time
social and individual process. The basic idea is that the process for an
individual to make a decision contain elements of both social influence and
individually obtained information. This is, since Janis's concept of
``groupthink''~\cite{janis}, a well-established idea in
psychology. Our model is a non-equilibrium, agent-based model for how
$N$ individuals each settle for one of two choices.
This dynamic model is grounded on two precepts: 
First, the information that an agent bases its decision on is a sum
of contributions from its social neighbors and direct information
sources.  Second,
when the information in favor of one decision exceeds a certain
threshold, the agent finalizes the decision. Furthermore, we assume the
underlying network to be static. To study the effect of the network
topology on the dynamics, we run it on different network models.

In the rest of the paper, we will first give a derivation and a
detailed description of the dynamic model and the models to generate the
underlying networks. Then we present our simulation results and finally
relate them to psychosocial phenomena.

\section{Preliminaries}

\subsection{Definition of the model}

We restrict ourselves to a binary decision process---a process
where the outcome of the decision can be good or bad, right or
wrong, correct or false. Without loss of generality, we let $+1$ represent the
``correct'' outcome. Consider a population of $N$ agents connected via
an underlying,
static, social network. The social network is represented by a graph
$G$ consisting of $N$ vertices, $V$; and $M$ edges, $E$. The
accumulated information used by an agent $i$ at time $t$ of the
decision making process is represented by $S_t(i)$. The
initial value $S_0(i)$, representing the direct information, is
picked randomly. The probability of $S_0(i)=+1$ is $p$, otherwise
$S_0(i)$ takes the value $-1$.

The social information-spreading process updates agents iteratively
until all agents have reached fixed states. The current choice of
an agent $i$ is $\mathrm{sgn}\: (S_t(i))$ where
\begin{equation}
  \mathrm{sgn}\:(x)=\left\{\begin{array}{rl} -1 & \mbox{if $x<0$}\\ 0 &
      \mbox{if $x=0$} \\  1 &
      \mbox{if $x>0$}\end{array}\right. .
\end{equation}
In our simulations $\mathrm{sgn}(S_t(i))=0$ represents the rare
situation the agent $i$ is completely undecided at time $t$.
If the information in favor of a particular decision is strong
enough, i.e.\ $|S_t(i)|> \theta$ for a threshold $\theta$, the
agent will finalize his/her decision, and $S_t(i)$ remain fix for the rest of the run.
The threshold $\theta$ is for simplicity considered to be the same for
every agent. A higher $\theta$ implies that the system needs more time
to converge to its decision. The information exchange between two
agents works as follows: Pick an active (not finalized) agent $i$ at random,
and a random neighbor $j$ of $i$ and let
\begin{equation}\label{eq:udi}
  S_{t+1}(i)=S_t(i)+\mathrm{sgn}\:S_t(j).
\end{equation}

To summarize the algorithm:
\begin{enumerate}
\item Pick $p$ and $\theta$.
\item Assign each $S_0(1),\cdots,S_0(N)$; $1$ with probability $p$, and
  $-1$ otherwise.
\item \label{item:pick} Pick a random active vertex $i$ and a random of $i$'s
  neighbors $j$.
\begin{enumerate}
 \item Let $S_{t+1}(i)=S_t(i)+\mathrm{sgn}\: S_t(j)$.
\item If $|S_{t+1}(i)|> \theta$, let agent $i$ be finalized.
\end{enumerate}
\item If there are unfinalized vertices, go to step \ref{item:pick}.
\end{enumerate}
Our model thus have two control parameters, $p$ and $\theta$.

\subsection{The model: considerations and possible extensions}

In a real situation the direct knowledge  may be obtained any time during a
group prediction process. We believe this does not change the
qualitative behavior of the model much. Assuming the direct
information is to the benefit of the agent, the sensible range of $p$ is
$[1/2,1]$. If $p=1/2$ the initial knowledge does
not guide agents towards a correct decision at all; if $p=1$ the
knowledge or preference influences the final decision strongly.

One also has to consider the relative influence of the agents in the
social network. As mentioned, in our model, one vertex $i$ and a
random neighbor of $i$'s is selected for the information transfer. This
means that, if the network has neutral degree-degree correlations,
then the
probability that information is obtained from a vertex with degree $k$ is
proportional to $k$. I.e., vertices with many connections are more
likely to influence others than vertices with few connections. This we believe
is a plausible situation: People with many social ties are more likely
to function as opinion-makers.

\subsection{Network models}

The model we present can be applied to any underlying network. In this
paper we will use three types of model networks:
The first model is random graphs~\cite{janson} constructed such that
$M$ edges are iteratively added between random pairs of vertices
such that no multiple edge or loop (self-edge) occurs. Such graphs have
a binomial degree distribution (which becomes Poissonian in the
$N\rightarrow\infty$ limit) and no other network structure. Thus
random graphs make a good starting point for investigating a dynamic
model such as ours.

Since many real-world networks have skewed degree distributions we also
use a model producing networks with a power-law degree
distribution. These networks, that we will refer to as ``scale-free
networks,'' are constructed by first assigning desired degrees $\tilde{k}_i$
from a power-law distribution $\mathrm{Prob} (k) \sim
k^{-\gamma}$.
Then we assign edges to random pairs $(i,j)$ of vertices as long as
the degree of a vertex $i$ is less than $\tilde{k}_i$. If $(i,j)$
already is an edge, or $i=j$ we do not add an edge.
This process is terminated when $M$ edges are added.

\section{Numerical results}

\begin{figure}
\center
  \resizebox*{\linewidth}{!}{\includegraphics{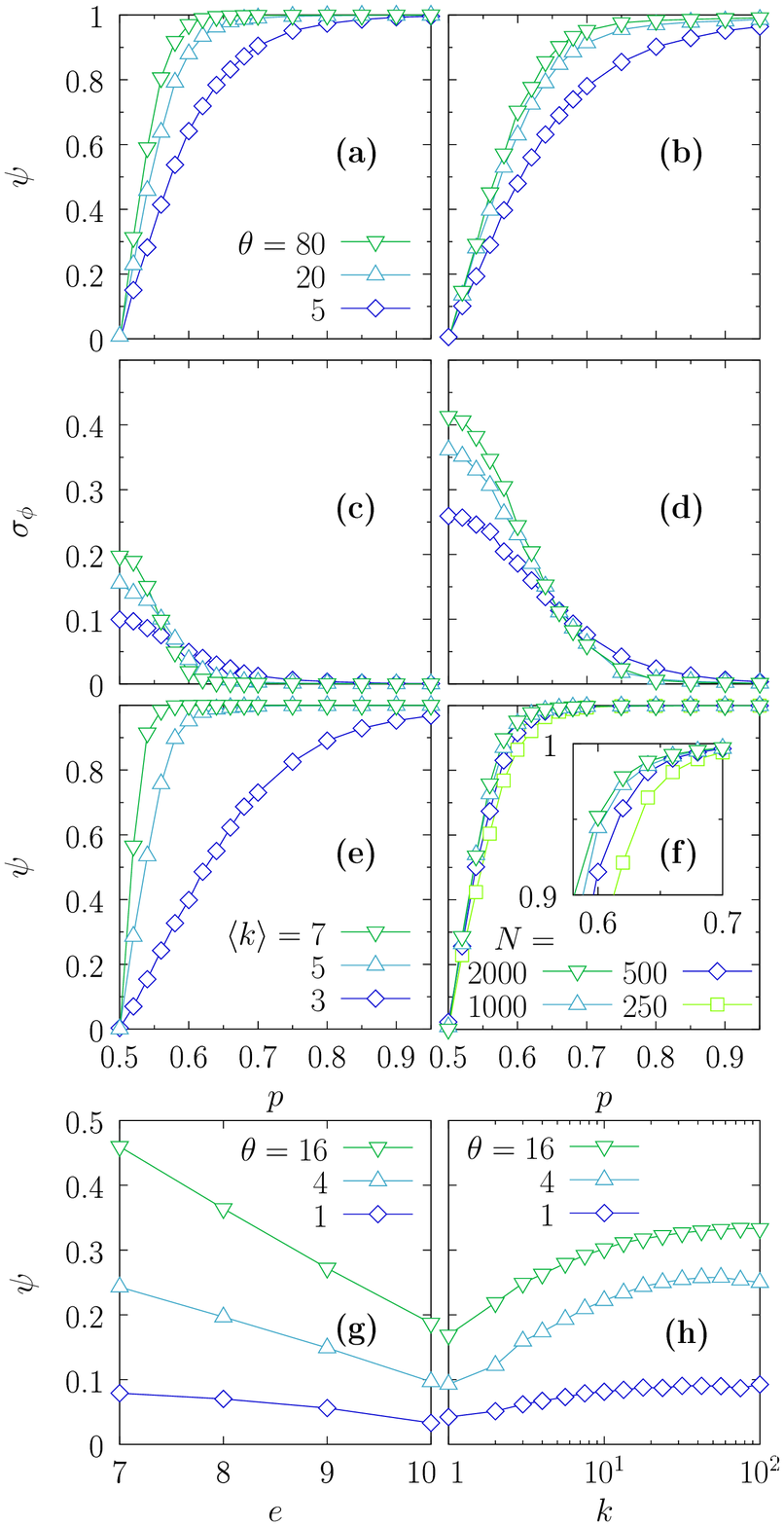}}
  \caption{
    Average relative increased performance $\psi$ as a function of information
    quality $p$ for (a) random graphs with different $\theta$-values,
    using  $N=1000$ and $\langle k\rangle = 5$, and (b) scale-free networks 
    with  $N=1000$, $\langle k\rangle = 5$ and $\gamma=2.2$.
    The standard deviaton $\sigma_{\phi}$ of the success rate $\phi$ for (c) random graphs and
    (d) scale-free networks.
    The average relative increased performance $\psi$, (e) as function of average 
    connectivity $\langle k\rangle$ for random graphs with
    $N=1000$ and $\theta=40$, and (f) as a function of system size $N$
    for random graphs with $\langle k\rangle = 5$ and $\theta=40$.
    Panel (g) shows the average relative increased performance $\psi$ as 
    a function of eccentricity for random graphs with
    $N=2000$, $\langle k\rangle =5$ and $p=0.53$.
    (h) displays $\psi$ as a function of degree for scale-free networks with
    $N=4000$, $\langle k\rangle =5$, $\gamma=2.2$ and $p=0.53$. The
    curves are logarithmically binned.
    Lines are guides for the eyes. All standard errors are smaller than the symbols.}
  \label{Fig1}
\end{figure}

\subsection{Success rate of the decision process}

Let $\phi$ be the fraction of the
population making the correct decision after the dynamics has
converged:
\begin{equation}\label{eq:phi}
  \phi = \frac{1}{2}+\frac{1}{2N}\sum_{i=1}^N \mathrm{sgn}\:S_\infty(i).
\end{equation}
We note that there are two types of qualitatively different behavior
that can emerge. Either the system benefit from the social
communication or it does not. To study this behavior we define the
order parameter
\begin{equation}\label{eq:psi}
  \psi=\frac{\phi-p}{1-p},
\end{equation}
i.e., the improvement of the decisions from the communication
relative to the theoretically largest improvement. If no improvement
of the decisions are made then $\psi=0$. If all agents decisions are
correct (i.e.\ the improvement is maximal) $\psi=1$. 

By construction $p=0.5$ is the symmetry value for which $\psi=0$. This means that as $p$
is increased from $p=0.5$, systems will on average benefit from the information diffusion.

In Fig.~\ref{Fig1}(a) we display $\psi$, for random networks. We see that
the decisions improve with, not only $p$, but also $\theta$. With a higher
decision threshold the agents need more information, and thus more iterations
to converge.  As more
information is integrated, the average agent can reach the correct
decision with higher probability. However, there is also a
possibility that a majority (or, indeed, any fraction) of the agents
end up with the wrong decision.

In Fig.~\ref{Fig1}(b) we display $\psi$, for scale-free networks. We see that,
as with the random networks, decisions improve with both $p$ and $\theta$.
Compared with random networks, the decisions for scale-free networks
improve less, on average, from communication. Qualitatively however,
the improvement from communication is similar to the case of random networks.

We note that the fluctuations of the success rate $\phi$ is larger for scale-free
than random networks as can be seen from~\ref{Fig1}(c) and (d).
These fluctuation, probably arising from the larger fluctuations in
degree, is a possible cause for the comparatively low $\psi$-values of
the scale-free networks (since, occasionally large hubs
will initially be assigned $S=-1$ and influence its many neighbors
towards an incorrect decision).

In Fig.~\ref{Fig1}(e) we display $\psi(p)$ for different average degrees $\langle k \rangle$
in random graphs. We see that $\psi$ increases for with increasing $k$, that is, systems of agents with less
restricted communication patterns will improve more from communications.

In Fig.~\ref{Fig1}(f) we display $\psi(p)$ for different system
sizes. As we can see, not much of a difference appears with different system size. At most a
smaller deviation is visible closer to the $p$ values for which $\psi$
approaches $1$.

How does the position of a vertex in the network affect its ability
to make correct decisions? To investigate this, we plot the average success of
vertices as a function of their eccentricity (the maximal distance
to any other vertex in the network) in Fig.~\ref{Fig1}(g) (for the
random network model).  A central (low eccentricity)
vertex has a higher probability of
making a correct decision than a more eccentric vertex. Since
centrality measures are usually strongly correlated~\cite{centr:keiko}
this suggests that being central in the information flow, in general, is beneficial
for the individual vertex.
Furthermore, in Fig.~\ref{Fig1}(h) we plot $\psi$ versus degree in a
scale-free network. As degree can be regarded as a local centrality
measure it is no surprise that these curves reveal a positive correlation.

\begin{figure}
\center
  \resizebox*{0.9\linewidth}{!}{\includegraphics{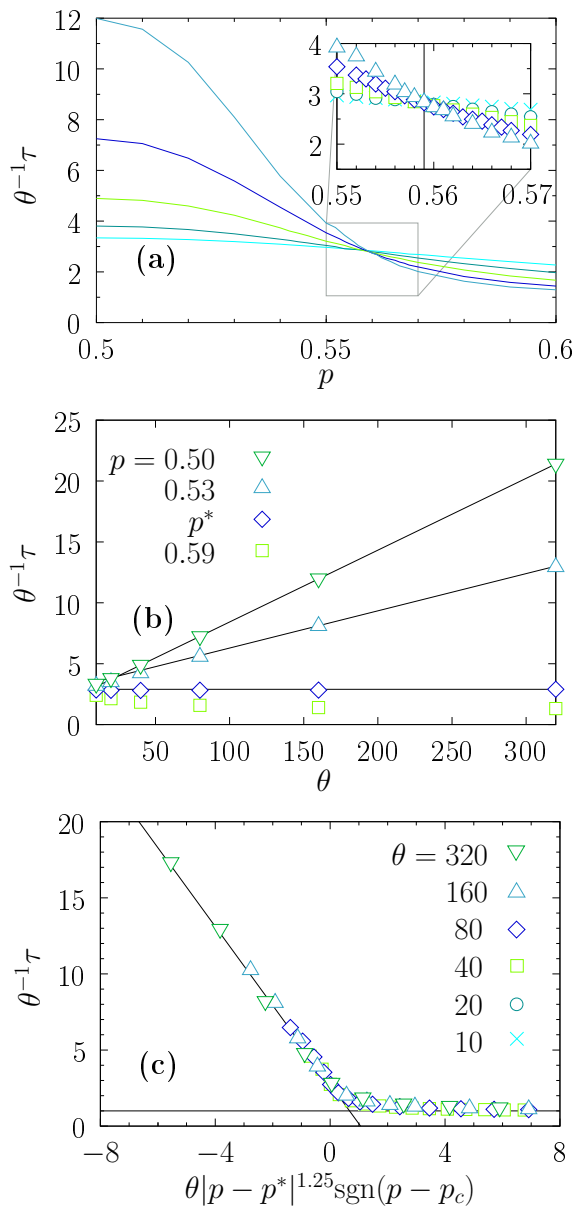}}
  \caption{Scaling analysis of $\tau$ in random graphs,
    where (a) is a crossing plot
used to determine $p^*=0.559\pm 0.001$. The $\theta$-values in (a)
are, from top to bottom in the left side of the figure (or bottom to
top in the right side): $160$, $80$, $40$, $20$, and $10$ (as
indicated by the legend in panel (c)).
(b) shows $\tau$'s $\theta$-scaling. Straight lines are fitted to the
    data points.
(c) displays a scaling-collapse plot verifying $F(x)$'s asymptotic behavior using
$\theta=320$, $160$, $80$ and $40$. For all networks $N=4000$ and
    $\langle k\rangle=5$. Standard errors are smaller than symbol
    sizes.}
  \label{Fig2}
\end{figure}

\subsection{Dynamic properties of the decision making}

Being a non equilibrium model, a natural quantity to study is the
average time for systems to reach the fixed state
(where all have finalized their decisions).
We call it the \textit{finalizing time} $\tau$.
To obtain an intensive time scale (i.e.\ independent of the system size $N$),
we let $t=t'/N$, where $t'$ is the simulation time for the system.
Since an agent has to update $S$ at least $\theta$ 
times to finalize its decision, we see that $\tau$ cannot scale slower
than linearly with $\theta$, that is if $\tau\sim\theta^z$,
we in all situations (all $p$) have $z\geq 1$.
In Fig.~\ref{Fig2}(a) we display $\theta^{-1}\tau$ for random graphs as
underlying topology. 
We observe a crossing point at $p$ close to $p^*=0.56$, below which $\theta^{-1}\tau$ is increasing with $\theta$ and for larger $p$, $\theta^{-1}\tau$ is decreasing with $\theta$.
Since $z\geq 1$ for all $p$ we should, in the limit of large $\theta$, have $z=1$ for $p>p^*$.
From Fig.~\ref{Fig2}(b) we observe for $p<p^*$, and for
large $\theta$, that $\theta^{-1}\tau$ seems to scale linearly with
$\theta$. In other words, for large $\theta$ and $p<p^*$, $\tau$ is
proportional to $\theta^2$.
To connect the different behaviors above and below $p^*$, we assume a scaling function of the form
\begin{equation}\label{eq:fss}
  \tau = \theta\,F\Bigl(\theta | p-p^* |^{\nu} \mathrm{sgn}(p-p^*))\Bigr).
\end{equation}
More specifically, to validate our scaling assumption $F(X)$ must be
constant in the
limit $X\rightarrow \infty$ (to verify $\tau \sim
\theta$ above $p^*$), and $F(X)\sim X$ in the limit $X\rightarrow
-\infty$ (to verify $\tau \sim \theta^2$ below $p^*$). In
Fig.~\ref{Fig2}(c) we display a collapse to verify the scaling
relation. The collapse is in agreement with our anticipated
scaling behavior and thus demonstrates the asymptotic behavior of
$F(X)$. To summarize, we have two distinct dynamical regions
$\tau\sim\theta^{z}$ with the critical indices $z=1$ and $z=2$ that
gives the following asymptotic scaling properties of the finalizing
time for the system:
\begin{equation}
  \tau\sim\left\{\begin{array}{ll}\theta & \mbox{if $p\geq p^*$}\\ 
          \theta^{2} & \mbox{if $p<p^*$}\end{array}\right..
\end{equation}

\begin{figure}
\center
  \resizebox*{0.9\linewidth}{!}{\includegraphics{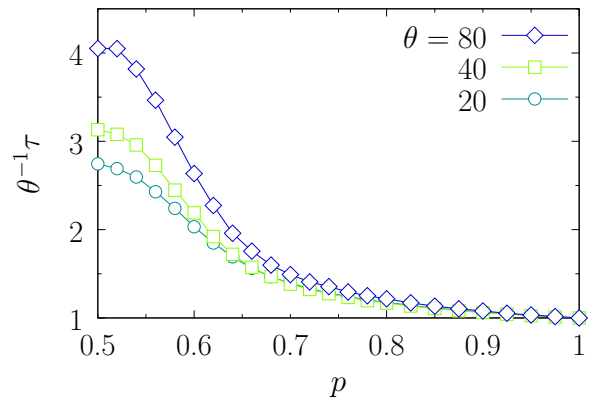}}
  \caption{The scaling of the finalization time for scale-free
    networks. In this figure, $N=4000$, $\langle k\rangle=5$ and
    $\gamma=2.2$.}
  \label{Fig3}
\end{figure}

In Fig.~\ref{Fig3} we investigate the $\theta$-dependence of $\tau$
for scale-free networks. Besides the trivial point at $p=1$, we do not
observe any crossing point of $\tau$. This suggests that we do not have
two distinct scaling regions for scale-free networks, rather $z$
seems to be continuously varying, $z=z(p)$, with
$1=z(1)<z(p<1)\leq p(1/2)\approx 1.4$. In comparison with random
graphs the scaling of $\tau$ for scale-free networks is faster for
$p<p^*$ but, in the large-$\theta$ limit, slower for $p>p^*$.

\begin{figure}
\center
  \resizebox*{\linewidth}{!}{\includegraphics{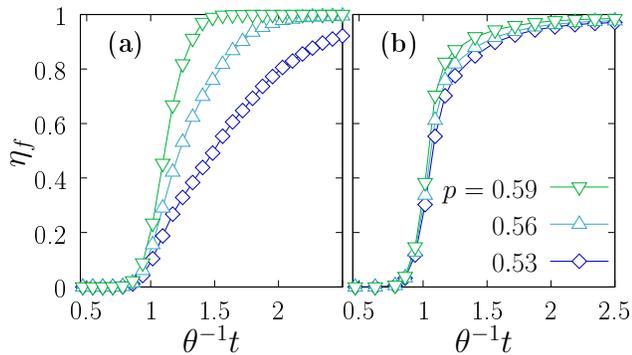}}
  \caption{The fraction of finalized vertices $\eta_f$ over time in (a)
    random graphs, $N=8000$, and (b) scale-free networks with $N=4000$ and
    $\gamma=2.2$. In all runs we have $\langle k\rangle=5$ and
    $\theta=80$. Lines are guides for the eyes. Standard errors are
    smaller than symbol sizes.}
  \label{Fig4}
\end{figure}

To get a closer view of the time evolution, we plot (in
Fig.~\ref{Fig4}(a)) the average number of finalized agents, $\eta_f$, as
function of time.
We choose three $p$-values ($0.53$, $0.56$ and $0.59$)---over, at, and
below the transition between the scaling regimes for random graphs. 
We plot the corresponding curves for scale-free networks in Fig.~\ref{Fig4}(b). 
The curves for $p=0.53$, $0.56$ and $p=0.59$ are similar,
as can be expected from Fig.~\ref{Fig3}. The initial increase of
$\eta_f$ is as fast for the curve of the scale-free networks as for
the $p=0.59$-curve of random networks, but the approach to $\eta_f=1$
at later times slows down considerably. Thus, even if the time a whole
population reaches finalization faster for random networks, the time
an intermediate fraction of the population have finalized their
decisions may be faster for the SF-networks.

\section{Summary and discussion}

We have proposed a dynamic model for collective decision
processes. A decision is a choice of one of two values---the
``correct'' value $+1$ and the ``incorrect'' $-1$.
We assume an individual's decision is affected by both directly
obtained information (with some degree of helpful content) and
information from social neighbors over social networks.
We find that the community will benefit from social interaction
whenever directly obtained information is the least helpful. 
We also observe that systems of agents with less
restricted communication patterns will improve more from sharing information.
Different global network topologies do not affect the
decision ability quantitatively very much. In other studies,
dynamical network systems---like spatial
games~\cite{abramson,ebelbornholdt:coevo,our:realpd}, disease
spreading~\cite{vesp:episf,andersonmay}, congestion sensitive
transport~\cite{holme:traffic}, and so on---are highly sensitive
to the structure of the underlying networks.
The local network structure makes a quantitative difference between the agents.
An agent that is central in the information flow has a significantly higher
chance of making the correct decision, than more peripheral agents.

Generally speaking, the population reach, on average, a decision that
is more likely to be correct than a random guess. There might however
be clusters of incorrect predictions, occasionally spanning a large
fraction of the population. In psychology this result, that collective
predictions may be irrational, has traditionally been explained by the
``groupthink'' concept of Irving Janis~\cite{janis}. In Janis's work
the driving force is that the members of a tight-knit group try to
conform to, what they believe, is the group's consensus, this creates
an unstable situation that may lead to an illogical conclusion. (For
modern works see Refs.~\cite{judith,zucker,gthnk:park} and further
references therein.) In our model, in contrast, the agents do not
actively try to avoid behaving aberrant; furthermore, the underlying
network need not be a fully connected network; yet sometimes the
majority of the population may predict falsely. In networks with
broader degree distributions this tendency is stronger---the large
fluctuation, gives large fluctuations in influence of the vertices and
obscures the system's ability to integrate the information of a large
fraction of the agents.
To epitomize, the dynamics of social information diffusion may, 
in rare cases, be enough to misguide a population and this is another
process that may lead to the same result as groupthink.

When we run our model on random graphs we find
two distinct dynamic regions of the convergence of the finalizing
time, $\tau\sim\theta^{z}$, with the critical indices $z=1$ ($p>p^*$)
and $z=2$ ($p<p^*$). In contrast, for scale-free networks we
find no distinct dynamic regions and the scaling is slower than
$\tau\sim\theta$ for all values $p<1$. 
In other words, if the average initial knowledge is high enough
($p>p^*$), the decision process on random graphs reach collective
decisions faster than on scale-free graphs. If, on the other hand,
the average knowledge is low ($p<p^*$) the decision making is faster
on scale-free networks. We have investigated the limit $\theta\to\infty$ for parameter values
such that the average distances $d$ in the network is significantly less
than $\theta$. This means that all the vertices of the network, in
principle, can influence all other vertices. A potentially different
case, which we leave for future studies, is if $\theta\ll d$ as $N$
grows. Since, in almost any random network, $d(N)$ is a very slowly
growing function (logarithmic or, in the case of SF networks, slower
still), it seems reasonable to assume that natural systems belong to the case we study.

Finally we note that there is a potential application of collective
decision models to computer science. In the theory of distributed
computing the \textit{Byzantine agreement} problem is to coordinate
concurrent processes where a number of the processes are
faulty~\cite{byzantine1,byzantine2}. In our context that would be to
make all vertices converge to the $+1$-prediction. We will not go into
details about algorithms to solve the Byzantine agreement problem, but
note that there is an area in common for mechanistic models of social
processes and algorithmic computer science. Also algorithms for
inference problems, such as belief propagation~\cite{frey:belief} or
models of associative memory~\cite{hertz} have partly common ground
with decision making models such as ours.

\acknowledgments
P.H. acknowledges financial support from the Wenner-Gren foundations
and the Swedish Foundation for Strategic Research.


\begin{thebibliography}{10}

\bibitem{schelling}
T.~C. Schelling,
  \textit{Micromotives and macrobehavior}
  (W.\ W.\ Norton \& Company, New York, 1978).

\bibitem{sznajd}
K. Sznajd-Weron and J. Sznajd,
 Int. J. Mod. Phys. C \textbf{11}, 1157 (2000).

\bibitem{deuff:opi}
G. Deffuant, F. Amblard, G. Weisbuch, and T. Faure,
  Journal of Artificial Societies and Social Simulation
  \textbf{5}, 1 (2002).

\bibitem{sood:vot}
V. Sood and S. Redner, Phys. Rev. Lett. \textbf{94},
  178701 (2005).

\bibitem{our:vot}
P. Holme and M.~E.~J. Newman, Phys. Rev. E \textbf{74}, 056108 (2006).

\bibitem{watts:fad}
D.~J. Watts, Proc. Natl. Acad. Sci. USA \textbf{99},
  5766 (2002).

\bibitem{our:fad2}
A. Gr\"{o}nlund and P. Holme,
  Advances in Complex Systems \textbf{8}, 261 (2005).

\bibitem{lindgrennordahl}
K. Lindgren and M.~G. Nordahl, Physica D \textbf{75}, 292 (1994).

\bibitem{ebelbornholdt:coevo}
H. Ebel and S. Bornholdt, Phys. Rev. E \textbf{66}, 056118 (2002).

\bibitem{our:realpd}
P. Holme, A. Trusina, B.~J. Kim, and P. Minnhagen, Phys. Rev. E \textbf{68}, 030901 (2003).

\bibitem{zucker}
S. Galam and J.-D. Zucker, Physica A \textbf{287}, 644 (2000).

\bibitem{fore:marsili}
P. Curty and M. Marsili, J. Stat. Mech., P03013 (2006).

\bibitem{janis}
I.~L. Janis, \textit{Victims of Groupthink: A psychological study of
  foreign-policy decisions and fiascoes} (Houghton
  Mifflin, Boston, 1972).

\bibitem{janson}
S. Janson, T. {\L}uczac, and A. Ruci\'{n}ski,
  \textit{Random Graphs} (Whiley, New York, 1999).

\bibitem{centr:keiko}
K. Nakao, Connections \textbf{13}, 10 (1990).

\bibitem{chung_lu:pnas}
F. Chung and L. Lu, Proc. Natl. Acad. Sci. USA \textbf{99}, 15879
(2002).

\bibitem{abramson}
G. Abramson and M. Kuperman,  Phys. Rev. E \textbf{63}, 030901 (2001).

\bibitem{vesp:episf}
R. Pastor-Satorras and A. Vespignani, Phys. Rev. Lett. \textbf{86}, 3200 (2001).

\bibitem{andersonmay}
R.~M. Anderson and R.~M. May,
  \textit{Infectious diseases in humans}
  (Oxford University Press, Oxford, 1992).

\bibitem{holme:traffic}
P. Holme, Advances in Complex Systems \textbf{6}, 163 (2003).

\bibitem{judith}
J.~M. Puncochar and P.~W. Fox,
  Journal of Educational Psychology \textbf{96}, 582 (2004).

\bibitem{gthnk:park}
W.-W. Park, Journal of Organizational Behavior \textbf{21}, 873 (2000).

\bibitem{byzantine1}
M. Pease, R. Shostak, and L. Lamport, J. ACM \textbf{27}, 228 (1980).

\bibitem{byzantine2}
L. Lamport, R. Shosta, and M. Pease, ACM
Trans. Prog. Lang. Syst. \textbf{4}, 382 (1982).

\bibitem{frey:belief}
B. Frey, ``Graphical Models for Machine Learning and Digital
Communication'' (MIT Press, Cambridge MA, 1998).

\bibitem{hertz}
J. Hertz, A Krogh and R. G. Palmer, \textit{Introduction to the theory
  of neural computation} (Addison-Wesley, Redwood City, 1991).

\end{thebibliography}
\end{document}